\documentstyle[aps,preprint]{revtex}
\tightenlines
\begin{document}
\def\be{\begin{equation}}
\def\en{\end{equation}}
\def\bq{\begin{eqnarray}}
\def\eq{\end{eqnarray}}
\def\noi{\noindent}
\def\bi{\bigskip}
\def\tr{{\rm tr}}

\title{Strong and electromagnetic contributions to the $U_A(1)$ anomaly 
and the $P^0\to\gamma\gamma$ decays}

\author{ M. Napsuciale and S. Rodriguez}

\address {Instituto de Fisica, Universidad de Guanajuato,\\
Lomas del Bosque 103, Fracc. Lomas del Campestre \\ 
37150, Leon, Guanajuato, Mexico}
%{\small\it Instituto de Fisica, Universidad de Guanajuato,\\
%AP E-143, 37150, Leon, Guanajuato, Mexico }}
\maketitle

\begin{abstract}
Using flavor basis we relate flavor axial anomalies to the mass matrix of 
pseudoscalar isoscalar fields in the context of a Linear Sigma Model which 
includes $U_A(1)$ symmetry breaking. We incorporate additional contributions to 
these anomalies due to external electromagnetic fields invoking 't Hooft's argument 
on anomaly matching 
and work out the predictions of this formalism for $\eta\to\gamma\gamma$ and 
$\eta^\prime\to\gamma\gamma$ decays. We show that the only effect of the $U_A(1)$ 
anomaly in these processes is in the formation of the $\eta$ and $\eta^\prime$ 
systems. From experimental data on these decays we extract the 
pseudoscalar mixing angle in flavor basis as  
$\phi_P\in\lbrack 38.4^\circ ,41.0^\circ \rbrack $.
\end{abstract}

\bi
%\begin{keyword}
%flavor symmetry \sep eta meson \sep instantons \sep OZI rule 
%\end{keyword}

\section{Introduction.}

The understanding of the mechanisms leading to the  mixing of pseudoscalar 
mesons is an important task in hadronic physics and this topic has been actively 
investigated during the last years \cite{todos}. From the OZI rule perspective 
this mixing is unusually large, in contrast with e.g. vector mesons which are close 
to the ideal mixing composition dictated by the OZI rule. 
Mixing of pseudoscalar mesons has been traditionally described in the Gell-Mann 
basis for $SU(3)$. In this framework, octet axial currents are conserved in the 
massless quark limit 
\be
\partial^\mu A^a_\mu ~\equiv~\partial^\mu \bar q \frac{\lambda^a}{2}
\gamma_\mu\gamma_5 q = 0,~~~~ a=1,...,8 ,
\en

\noi whereas the singlet axial current has a non-vanishing 
divergence due to the gluonic ABJ triangular anomaly
\be
\partial^\mu A^0_\mu ~=~ \frac{1}{\sqrt{6}} \frac{n_f \alpha_s}{4\pi}
G^a_{\alpha\beta}\tilde G^{a\alpha\beta} 
\label{divsinglet}
\en
where $\tilde G^{a\alpha\beta}\equiv\frac{1}{2}
\varepsilon^{\alpha\beta\mu\nu}G^a_{\mu\nu}$. This has far-reaching 
consequences for  hadronic systems, in particular to those sharing 
the quantum numbers of the operators on the left of Eq.(\ref{divsinglet}). 
These relations must be modified in the presence of external vector fields to account
 for additional contributions coming from the coupling of fermions to these fields.
The typical example of modifications to these relations is the case of 
massless QCD coupled to external electromagnetic fields. In this case, e.g. 
the divergence of $A^3_\mu$ gets a contribution from the ABJ photon anomaly
\be
\partial^\mu A^3_\mu = \frac{\alpha}{4\pi}F_{\alpha\beta}\tilde F^{\alpha\beta}
\en
\noi where $F_{\alpha\beta}$ stands for the electromagnetic strength 
field tensor. This result, and the characterization of the matrix element of 
$A^3_\mu$ between the vacuum and two photons on the basis of Lorentz covariance, 
gauge invariance, parity etc. lead to the existence of a 
singularity at $q^2=0$ in this matrix element \cite{Shifman:1991zk}. 
In the confining 
theory, is meaningless to speak about the quarks as physical degrees of freedom
and the only explanation for this singularity invokes 't Hooft's consistency 
condition \cite{'tHooft:1979bh}, i.e. that singular contributions in 
$\langle 0|A^3_\mu |\gamma\gamma\rangle $ at the 
level of quarks and at the level of hadrons must match. This leads to a massless
hadronic excitation (the pion) which couples to $A^3_\mu$ and to two photons. 
This mechanism successfully describe the $\pi^0\to\gamma\gamma$ decay which in 
the absence of the anomaly would be forbidden ( actually of a lower power in $m_\pi$
 \cite{Weinberg:1996kr}) which is inconsistent with experimental results.

In principle, one could try a similar calculation for the $\eta \to\gamma\gamma $ and 
 $\eta' \to\gamma\gamma $ decays. However, here we encounter the problem of how 
to quantify  the effects of the strong contribution to the $U_A(1)$ anomaly 
in Eq.(\ref{divsinglet}) at the hadron level and how these effects influence  the 
mixing of pseudoscalar isoscalar fields.

In the conventional singlet-octet basis, the octet current gets no 
contributions from the gluon 
anomaly and a similar procedure to the case of the pion can be used to estimate 
the  decay amplitude for $\eta_8 \to \gamma\gamma$, although the extrapolation from 
$q^2=0\to m^2_{\eta_8}$ is more severe in this case . Relating this amplitude 
to the $\eta\to \gamma\gamma $ decay in principle requires a careful analysis of the 
mechanisms 
for mixing of pseudoscalars which nevertheless, in this case, seems to be 
quantitatively not so 
relevant due to the experimental fact that the naive mixing angle is small in 
this basis. In the case of 
the singlet we run in trouble due to the strong contribution to the singlet 
axial anomaly. Relating the physical
amplitudes to the singlet-octet ones is more problematic in this case and 
definitively requires to clarify the role of the axial anomaly in the mixing of 
pseudoscalars (or stated differently, to quantify the gluon content of pseudoscalar 
mesons).

The mixing of the pseudoscalar isoscalar fields can also be formulated in 
flavor basis $\lbrace \eta_{\rm ns}, \eta_{\rm s}\rbrace $. The most general form 
of the mass Lagrangian is
\be 
 {\mathcal L}_{mass} =-\frac{1}{2} (m^2_{\eta_{\rm ns}} \eta^2_{\rm ns}+ 
 m^2_{\eta_{\rm s}} \eta^2_{\rm s} + 
 2m^2_{\rm s-ns} \eta_{\rm s} \eta_{\rm ns}).\label{lmix}
\en
\noi where the last term account for the OZI rule violating 
$\eta_{\rm ns}-\eta_{\rm s} $ transitions. The precise origin of this term 
is still unclear. The standard chiral expansion uses the octet and singlet 
fields, hence, whatever the mechanism for mixing of flavor fields be, it is 
considered from the very start in this formalism. 

In the model-independent formulation of mixing in Eq.(\ref{lmix}) we can relate 
the mixing angle to the non-diagonal term $m^2_{\rm s-ns}$. The extraction of 
the pseudoscalar mixing angle following this procedure requires the precise 
quantification of all the mechanisms contributing to meson masses which is a 
difficult task. In \cite{Napsuciale:2001sm,Napsuciale:2001jv} it was shown that 
the $U_A(1)$ anomaly 
gives a sizeable contribution to $m^2_{\rm s-ns}$ via its coupling to the 
spontaneous breaking of chiral symmetry. Assuming that this is the only mechanism 
for mixing of flavor fields, a pseudoscalar mixing angle is obtained consistent 
with Gell-Mann's $SU(3)$ symmetry when pseudoscalar meson masses are used as input to 
fix the values of the free parameters of the model. 
Although interesting from the conceptual point of view, this mechanism does not 
account for the experimentally measured mixing angle which is close to, but 
definitively different from the $SU(3)$ value. The precise description of 
the mixing angle calls for considering further mechanisms among which Lipkin's 
loops cancellation \cite{Lipkin:1996ny} looks appealing as pointed out in 
\cite{Napsuciale:2001jv}. 
Another possibility for the extraction of this angle is to consider 
processes where this angle be involved but  sensitivity to pseudoscalar meson 
masses be reduced. 

Photonic decays of pseudoscalar mesons are  appropriate to this end  
since, as we shall see below, in this case strong contributions to the $U_A(1)$ 
anomaly can play a role in the conformation of physical pseudoscalar mesons only 
and effects due meson masses on the corresponding widths are softened.  

In this work we relate breaking of axial symmetry in the isoscalar channels to 
the pseudoscalar mass matrix in Eq.(\ref{lmix}), within a Linear Sigma Model 
which incorporates $U_A(1)$ symmetry breaking. Contributions of external 
electromagnetic fields to the breaking of axial symmetry in the isoscalar channels 
are introduced using 't Hooft's 
anomaly matching condition. We calculate $P^0\to\gamma\gamma$ decays in this 
framework and extract the pseudoscalar mixing angle from experimental data on these 
processes.   
 
In order to state notation we briefly review the model in the next section and 
work out its predictions for the isoscalar weak decay constants 
and the anomalies in the isoscalar axial currents. In section III we introduce 
effects of external 
electromagnetic fields and work out the predictions of this formalism for the 
$P^0\to\gamma\gamma$ decays. In section IV we give our conclusions.

\section{Axial anomalies and the mass matrix of isoscalar pseudoscalar fields.}

\subsection{The model.}

The  chirally symmetric $ \lbrack U(3)_L\otimes U(3)_R\rbrack $ meson Lagrangian 
\cite{Napsuciale:2001sm,Napsuciale:2001jv,Napsuciale:1998ip,Tornqvist:1999tn,'tHooft:1999jc}, describes 
a scalar and a pseudoscalar nonet, in turn denoted by $(\sigma_i)$ and $(P_i)$,
\begin{equation}
{\mathcal L}= {\mathcal L}_{sym}  + {\mathcal L}_{U_A(1)} + {\mathcal L}_{SB}\;.
 \label{lsm}
\end{equation}
Here, 
\begin{eqnarray}\nonumber
 {\mathcal L}_{sym}&=&\frac{1}{2}\tr\left[(\partial_\mu M)
 (\partial^\mu M^{\dagger}) \right]  -\frac{\mu^2}{2} 
 X (\sigma ,P)\\
 && \mbox{} -\frac{\lambda}{4} Y(\sigma ,P)
 -\frac{\lambda^\prime}{4} X^2 (\sigma , P) \; , 
 \label{symmlag}
\end{eqnarray}
$M=\sigma + i P$, and $X,Y$ stand in turn for the left-right
symmetric traces 
\be
 X(\sigma,P)=\tr\left[ M M^{\dagger}\right],~~  
 Y(\sigma,P)=\tr\left[ (M M^{\dagger})^2\right ]
\en

The pseudoscalar and
scalar matrix fields $P$ and $\sigma $ are written in terms of a specific
basis spanned by seven of the standard Gell-Mann matrices, namely $\lambda_i
~(i=1,\ldots , 7)$,  and by two non-standard matrices
$\lambda_{\rm{ns}}$=diag(1,1,0), and $\lambda_{\rm{s}}$ = $\sqrt{2}$
diag(0,0,1), respectively.  The decomposition obtained in this way reads
$P\equiv \frac{1}{\sqrt{2}} \lambda_i P_i$ with 
$i=ns,s,1,\ldots,7$ and similarly
for the scalar field. 
The instanton-induced interaction in
(\ref{lsm}) is 
\begin{equation}
 {\mathcal L}_{U_A(1)} = - \beta \left( \mbox{det}(M)+\mbox{det} (M^{\dagger }) 
\right). \label{lins} 
\end{equation}

It stands for the bosonization of 't Hooft's effective quark-quark 
interaction which has a determinant structure in flavor space 
~\cite{'tHooft:1976fv,'tHooft:1986nc}.
Finally, there is the standard quark mass term
\begin{equation}
 {\mathcal L}_{SB} =  \tr \left[ c\sigma \right]= 
\tr \left[ \frac{b_0}{\sqrt{2}}{\mathcal M}_q(M+M^{\dagger}) \right]\label{lbreak}
\end{equation}
which breaks the left-right symmetry explicitly.
The $c$ matrix is spanned by the same basis 
$c\equiv \frac{1}{\sqrt{2}} \lambda_i c_i$,
where the nine expansion coefficients $c_i$ are independent 
constants. It is related to the
quark mass matrix by $c=\sqrt{2}b_0{\mathcal M}_q$ and has  
$\frac{c_{\rm{ns}}}{\sqrt{2}}= \sqrt{2}\hat m b_0$ and $c_{\rm{s}}=\sqrt{2} m_s 
b_0$ as the only non-vanishing entries. Here, $b_0$ is an unknown parameter with 
dimensions of squared mass. We work in the exact isospin limit,
$\hat m =m_u=m_d$ in the following. 
The linear $\sigma$  term in Eq.\,(\ref{lbreak}) induces 
$\sigma$-vacuum transitions which supply the scalar
fields  with non-zero vacuum expectation values (v.e.v) 
(hereafter denoted by $\langle \cdots\rangle $). 
To simplify notations, let us re-denote 
$\langle \sigma \rangle $ by $V$ with $V= $diag ($a,a,b$), where  
$a$ and $b$ in turn denote the vacuum expectation values
of the strange and non-strange quarkonium, respectively,
\begin{equation}
 a=\frac{1}{\sqrt{2}}\langle \sigma_{\rm{ns}}\rangle\;, \qquad
 b=\langle \sigma_{\rm{s}}\rangle \;. \label{vev}
\end{equation}
We now shift, as usual, the old $\sigma $ field to a  new scalar
field
$S=\sigma -V$ such that $\langle S\rangle=0$. In this way, 
new  mass terms, three-meson interactions,
and a linear term are generated. All these terms are affected -- via the
't Hooft determinant 
by  the $U_A(1)$ anomaly which get coupled 
to the v.e.v' s of the scalar fields by the spontaneous breaking 
of chiral symmetry. The consequence of
all these effects is the breaking of the original symmetry 
down to $SU(2)_I$ isospin.
The masses of the seven unmixed pseudoscalar corresponding to the 
original Gell-Mann matrices
$\lambda_i$ ($i=1,\ldots,7$), namely 
the isovector pseudoscalar ($\pi$) mesons as well as 
the two  isodoublets of 
pseudoscalar  ($K$) mesons, are obtained as 
\cite{Napsuciale:2001sm,Napsuciale:2001jv,Napsuciale:1998ip,Tornqvist:1999tn,'tHooft:1999jc} 
\be
 m^2_\pi =\xi+2\beta b+\lambda a^2 ,~~~
 m^2_K =\xi+2\beta a+\lambda (a^2\mbox{$-$}ab\mbox{+}b^2);
 \label{unmixedmasses}
\en
where we used the convenient short--hand notation 
$\xi \equiv \mu^2 +\lambda^\prime(2a^2+b^2)$. 
The elimination of the linear terms 
imposes the following constraints on the explicit-symmetry-breaking terms
$c_{\rm{ns}}$, and $c_{\rm{s}}$:
\begin{equation}
 c_{\rm{ns}} = \sqrt{2} a m^2_\pi\;,\qquad
 c_{\rm{s}} + \frac{c_{\rm{ns}}}{\sqrt{2}}= (a+b) m_K^2 \; .
 \label{c-eq}
\end{equation}
\noi or in term of the quark masses
\be
am^2_\pi=\sqrt{2}\hat m b_0,\qquad (a+b)m^2_K=\sqrt{2}(\hat m+m_s)b_0.\label{mpimq}
\en

In Ref.\cite{Napsuciale:1998ip} the  PCAC relations for the pion and kaon field are
discussed. These relations yield
\begin{equation}
 f_\pi =\sqrt{2} a \;, \qquad
 f_K = \frac{1}{\sqrt{2}} (a+b)\;,\label{fkfpi}
\end{equation}
\noi which when used in (\ref{mpimq}) yields
\be
f_\pi m^2_\pi = 2\hat mb_0,\qquad f_K m^2_K= (\hat m+m_s)b_0.
\en

The mass term of the Lagrangian involving the mixed isoscalar
pseudoscalar  fields, which correspond
to the $\lambda_{\rm ns}$ and $\lambda_{\rm s}$
matrices, has the  structure of the mass Lagrangian in Eq.(\ref{lmix}) with the 
specific values
\be
m^2_{\eta_{\rm{ns}}} =\xi-2\beta b + \lambda a^2 ,~~~ 
  m^2_{\eta_{\rm {s}}} =\xi + \lambda b^2 ,~~~~   
 m^2_{\rm {s-ns}} = -2\sqrt{2}\beta a   
 \label{snsmixing}
\en
Here, $m_{\eta_{\rm s}}$ and $m_{\eta_{\rm{ns}}}$ are  the masses of the 
strange and non-strange pseudoscalar quarkonia respectively, while $m^2_{\rm s-ns}$, 
denotes the  transition mass-matrix elements of the strange--non-strange 
pseudoscalar quarkonia, which  in this model  is due to
the interplay between 't Hooft interaction and  the spontaneous breakdown of chiral 
symmetry. 

This mass Lagrangian can be diagonalized by rotating to the physical basis
\be
 \left( \begin{array}{c} \eta \\ \eta' \end{array}\right)=
R(\phi_P) \left( \begin{array}{c} \eta_{\rm ns} \\ \eta_{\rm s} \\
\end{array}\right)~~~{\rm with}~~~
R(\phi_P)=\left( \begin{array}{lclc} {\rm cos} \phi_P,  &  -{\rm sin} \phi_P \\
 {\rm sin}\phi_P, & {\rm cos}\phi_P \\
\end{array} \right), \label{mixrel}
\en
\noi such that the diagonal physical mass matrix $M_D \equiv Diag~( m^2_{\eta}, 
m^2_{\eta'})$ is related to the same matrix in the flavor fields 
representation $M_F$ as 
\be
M_D=R(\phi_P)M_F R^{\dagger}(\phi_P) \qquad {\rm where}\qquad 
M_F=\left( \begin{array}{lclc} m^2_{\eta_{\rm ns}},  &  m^2_{\rm s-ns} \\
 m^2_{\rm s-ns}, & m^2_{\eta_{\rm s}} \\
\end{array} \right). \label{massrels}
\en

\subsection{Flavor weak decay constants}

The flavor weak decay constants $f_{\rm ns},~f_{\rm s}$ are defined by:
\be
\langle 0|A^{\rm{ns}}_{\mu}(0)|\eta^{\rm{ns}}(q)\rangle
=if_{{\rm{ns}} }q_\mu , \qquad  
\langle 0|A^{\rm{s}}_{\mu}(0)|\eta^{\rm{s}}(q)\rangle
=if_{{\rm{s}} }q_{\mu}.  \label{wdc}
\en

In the literature we also find weak decay constants related to the 
following matrix elements \cite{Feldmann:2000uf}
\be
\begin{array}{c}
\langle 0|A^{\rm{ns}}_{\mu}(0)|\eta(q)\rangle
=if_{\eta}^{{\rm{ns}} }q_\mu , \qquad \langle 0|A^{\rm{ns}}_{\mu}(0)|\eta^\prime (q)
\rangle=i f_{\eta^\prime}^{{\rm{ns}} }q_\mu , \\
\langle 0|A^{\rm{s}}_{\mu}(0)|\eta(q)\rangle
=if_{\eta}^{{\rm{s}} }q_\mu , \qquad 
\langle 0|A^{\rm{s}}_{\mu}(0)|\eta^\prime (q)\rangle
=if_{\eta^\prime}^{{\rm{s}} }q_\mu .
\end{array} \label{cwdc}
\en   
Let us analyze the predictions of the model for the divergences of the isoscalar 
currents and their implications for the weak decay constants.
Under the axial transformations
\be
\delta M = -\frac{i}{\sqrt{2}}\lbrace\varepsilon ,M\rbrace ~~~,~~~
\delta M^\dagger = \frac{i}{\sqrt{2}}\lbrace\varepsilon ,M^\dagger \rbrace,
\en
\noi the Lagrangian in Eq.(\ref{lsm}) is no longer invariant due to the 
breaking terms. A calculation of the divergences of the strange and 
non-strange  axial currents in the model yields:
\be
\partial^\mu A^{\rm{ns}}_{\mu}  = c_{\rm{ns}} \eta_{\rm ns} + 2\beta W ,
\qquad \partial^\mu A^{\rm{s}}_{\mu}  =\sqrt{2} c_{\rm{s}} \eta_{\rm s} + 
\sqrt{2}\beta W \label{divcurr}
\en
\noi where $W$ stands for the contribution coming from t''Hooft interaction and 
contains trilinear, bilinear and linear terms in the fields. Explicitly
\be
W= i( det(M)- det M^{\dagger} )= -2\sqrt{2}ab\eta_{\rm ns}-2a^2\eta_{\rm s} + 
bilinear ~+ ~tril. \label{W}
\en
The bilinear and trilinear terms in Eq.(\ref{W}) give vanishing contributions 
at tree level to the quantities to be considered here, hence they will be 
dropped in the following . Inserting (\ref{divcurr},
\ref{W}) in Eqs.(\ref{wdc}) we obtain    
\be
f_{\rm ns} m^2_{\eta_{\rm ns}}= c_{\rm ns} -4\sqrt{2} \beta ab, \qquad 
f_{\rm s} m^2_{\eta_{\rm s}}=\sqrt{2} c_{\rm s}- 2\sqrt{2}\beta a^2. \label{npcac}
\en

Also from Eqs(\ref{unmixedmasses},\ref{c-eq},\ref{snsmixing}) we obtain the 
following relations 

\be
m^2_{\eta_{\rm ns}}-m^2_\pi = -4\beta b \qquad b m^2_{\eta_{\rm s}}+2\beta a^2= 
c_{\rm s}, \label{mnsmpi}
\en
which when inserted in Eq.(\ref{npcac}) predict \cite{Napsuciale:2001jv}
\be
f_{\rm ns} =\sqrt{2}a=f_\pi ~~~~f_{\rm s} = \sqrt{2} b = 2f_K -f_\pi. \label{fnsfs}
\en

On the other hand, inserting Eqs.(\ref{W},\ref{npcac}) in Eqs.(\ref{divcurr}) we 
obtain

\be
\partial^\mu A^{\rm ns}_\mu = f_{\rm ns}m^2_{\eta_{\rm ns}}\eta_{\rm ns} 
-4\beta a^2 \eta_{\rm s}, ~~~\partial^\mu A^{\rm s}_\mu = f_{\rm s}m^2_{\eta_{\rm s}}
\eta_{\rm s} -4\beta ab \eta_{\rm ns}. \label{divflavcurr}
\en

The last terms in the r.h.s of the previous two Eqs. are entirely 
due to the coupling of the $U_A(1)$ anomaly to the v.e.v.'s of scalars. They are 
a manifestation, at the hadron level, of the gluon ABJ anomaly in QCD which we 
assume here as dominated by 
instantons which generate 't Hooft interaction.  Eqs.(\ref{divflavcurr}) can be 
rewritten in a symmetric form 
\be
\begin{array}{c}
\frac{1}{f_{\rm ns}}\partial^\mu A^{\rm ns}_\mu = m^2_{\rm ns}\eta_{\rm ns} 
+m^2_{\rm s-ns} \eta_{\rm s}, \\ 
\frac{1}{f_{\rm s}}\partial^\mu A^{\rm s}_\mu = m^2_{\rm s-ns} \eta_{\rm ns}+
 m^2_{\rm s}\eta_{\rm s},
\end{array} \label{fundrel}
\en
\noi which makes explicit the relation between the  mass matrix of isoscalar 
pseudoscalars and the divergence of isosinglet axial currents . It is interesting to make also 
explicit terms driven by the quark masses and those induced by the $U_A(1)$  anomaly 
which are not related to quark masses. 
Using relations (\ref{c-eq},\ref{mnsmpi}), Eqs.(\ref{divflavcurr}) can also be  
rewritten  as 
 \be
\begin{array}{c}
\partial^\mu A^{\rm ns}_\mu = (2b_0 \hat m -2\sqrt{2}\beta f_{\rm ns} 
f_{\rm s}) \eta_{\rm ns} 
-2\beta f^2_{\rm ns} \eta_{\rm s}, \\
\partial^\mu A^{\rm s}_\mu =  (2b_0 m_s -\sqrt{2}\beta f^2_{\rm ns})
\eta_{\rm s} -2\beta f_{\rm ns} f_{\rm s} \eta_{\rm ns}.
\end{array}\label{fundrel1}
\en
\noi which makes transparent that in the absence of the $U_A(1)$ symmetry 
breaking ($\beta =0$) the divergences of flavor currents are driven by quark masses 
and flavor isoscalar pseudoscalar fields are pseudo-Goldstone bosons.
Eq.(\ref{fundrel}) is the fundamental relation which we will exploit below 
in the description of the two photon decay of isoscalar pseudoscalar mesons.
Before this, let us  make two remarks on the consequences of Eq.(\ref{fundrel}). 
The first one concerns  the weak decay constants.
Inserting (\ref{fundrel}) in the divergence of Eq.(\ref{cwdc}) and using 
Eq.(\ref{mixrel}) we obtain
\be
\begin{array}{c}
f^{\rm ns}_{\eta}~m^2_\eta =f_{\rm ns} \left(m^2_{\rm ns} \cos\phi_P -m^2_{\rm s-ns}
\sin\phi_P\right),\\ 
f^{\rm s}_{\eta}~m^2_\eta =f_{\rm s} \left(  m^2_{\rm s-ns}\cos\phi_P  - m^2_{\rm s} 
\sin\phi_P\right), \\
f^{\rm ns}_{\eta^\prime}~m^2_{\eta^\prime} =f_{\rm ns} \left( m^2_{\rm s-ns}
\cos\phi_P + m^2_{\rm ns} \sin\phi_P \right),\\ 
f^{\rm s}_{\eta^\prime}~m^2_{\eta^\prime} =f_{\rm s} \left(  m^2_{\rm s-ns}
\sin\phi_P  + m^2_{\rm s} \cos\phi_P\right).
\end{array}\label{frels}
\en

On the other hand, from Eqs. (\ref{massrels}) we get
\be
\begin{array}{c}
m^2_\eta \cos\phi_P = m^2_{\rm ns} \cos\phi_P -m^2_{\rm s-ns}\sin\phi_P ,\\ 
- m^2_\eta \sin\phi_P =  m^2_{\rm s-ns}\cos\phi_P  - m^2_{\rm s} \sin\phi_P , \\ 
m^2_{\eta^\prime} \sin\phi_P = m^2_{\rm s-ns}\cos\phi_P + m^2_{\rm ns} \sin\phi_P , 
\\
m^2_{\eta^\prime} \cos\phi_P = m^2_{\rm s-ns}
\sin\phi_P  + m^2_{\rm s} \cos\phi_P.
\end{array}
\en

\noi which when inserted in Eq.(\ref{frels}) yields
\be
\left(\begin{array}{cc}
f^{\rm ns}_{\eta} & f^{\rm s}_{\eta} \\
f^{\rm ns}_{\eta^\prime} & f^{\rm s}_{\eta^\prime}
\end{array}\right)=
\left( \begin{array}{cc}
\cos\phi_P &-\sin\phi_P \\
\sin\phi_P & \cos\phi_P 
\end{array}\right)
\left(\begin{array}{cc}
f_{\rm ns} & 0 \\
0 & f_{\rm s}
\end{array} \right). \label{frels1}
\en

This relation was {\it postulated} in \cite{Feldmann:1998vh} and taken as the basic 
assumption in the analysis of pseudoscalar mixing and weak decay constants.

The second remark concerns the structure of the mass matrix and the contributions 
coming from quark mass terms. From Eqs.(\ref{fundrel1}) it is possible 
to write the mass matrix in the flavor basis as
\be
{\mathcal M}_F=\left( \begin{array}{cc}
m^2_{\rm qq}+2\alpha^2 &\sqrt{2}\alpha^2 y \\
\sqrt{2}\alpha^2 y & m^2_{\rm ss}+\alpha^2 y^2 
\end{array}\right)
\en
\noi where 
\be
m^2_{\rm qq}=\frac{2\hat m}{f_{\rm ns}}b_0,\qquad 
m^2_{\rm ss}= \frac{2 m_s}{f_{\rm s}} b_0,\qquad 
\alpha^2 =-2\beta b ,\qquad y=\frac{f_{\rm ns}}{f_{\rm s}}=\frac{a}{b}.
\en

It was shown in \cite{Feldmann:1998vh} that this result can be derived directly 
from QCD {\it if } we assume Eq.(\ref{frels1}). In terms of QCD quantities we identify 

\be
b_0=\langle 0|\bar ui\gamma_5 u+\bar d i\gamma_5 d |\eta_{\rm ns}\rangle ,\qquad 
\alpha^2 = \frac{1}{\sqrt{2}f_{\rm ns}}\langle 0| 
\frac{\alpha_s}{4\pi}G\tilde G |\eta_{\rm ns}\rangle .
\en

\bi

\section{Electromagnetic contributions to $U_A(1)$-symmetry breaking and 
$P^0\to\gamma\gamma $ decays.}

In the presence of external electromagnetic fields, relations(\ref{fundrel}) 
must be 
modified to account for the ABJ terms due to the photons. This modification 
can be obtained using   't Hooft's argument on the 
matching of the anomalies  \cite{'tHooft:1979bh} to translate 
exactly the same form of the anomaly as calculated at the level of the 
fundamental theory (QCD) to the composite  theory. For purposes of 
comparison below we include also the corresponding  relation for pions

\bq \nonumber
\partial^\mu A^{3}_\mu &=& f_{\pi} m^2_{\pi}\pi^0+\frac{\alpha}{4\pi}N_c 
D^{3}\varepsilon_{\mu\nu\rho\sigma }F^{\mu\nu} F^{\rho\sigma},  \\ 
\partial^\mu A^{\rm ns}_\mu &=& f_{\rm ns}\left( m^2_{\rm ns}\eta_{\rm ns} 
+m^2_{\rm s-ns} \eta_{\rm s}\right) +\frac{\alpha}{4\pi}N_c D^{\rm ns}
\varepsilon_{\mu\nu\rho\sigma }F^{\mu\nu} F^{\rho\sigma},\\ \nonumber 
\partial^\mu A^{\rm s}_\mu &=& f_{\rm s}\left( m^2_{\rm s}
\eta_{\rm s} +m^2_{\rm s-ns} \eta_{\rm ns}\right) +\frac{\alpha}
{4\pi}N_c D^{\rm s}
\varepsilon_{\mu\nu\rho\sigma }F^{\mu\nu} F^{\rho\sigma}.
\eq
\noi Here, $D$ denote the group 
factors $D^{\chi} = tr( \{Q,Q\}\frac{\lambda^{\chi}}{2})$ with 
$\chi=3,{\rm ns,s}$ and $Q$ stands the quark charge matrix. 
The last two relations can be written in terms of physical 
pseudoscalar fields and in  compact matrix form read 
\be
\left( \begin{array}{c} \frac{1}{f_{\rm ns}}
\partial A^{\rm ns}  \\ 
\frac{1}{f_{\rm s}}\partial A^{\rm s} 
 \end{array} \right) 
=R^{\dagger}(\phi_P)M_D\left(\begin{array}{c}\eta \\ \eta^\prime \end{array}\right) +
\left(\begin{array}{c}\frac{1}{f_{\rm ns}}D^{\rm ns} \\ \frac{1}{f_{\rm s}}D^{\rm s} 
\end{array}\right)\xi
\label{divfis}
\en
\noi where $\xi\equiv \frac{\alpha}{4\pi}N_c 
\varepsilon_{\mu\nu\rho\sigma }F^{\mu\nu} F^{\rho\sigma}$, $M_D$ denotes the 
diagonal mass matrix which is related to the same matrix in the flavor basis 
by Eq.(\ref{massrels}).

The three-point function for an axial and two electromagnetic currents 
\be
T^a_{\mu\nu\lambda}(k_1,k_2,q)\equiv i\int d^4xd^4yd^4z e^{i(k_1\cdot 
x+k_2\cdot y
-q\cdot z)}\langle 0|Tj_\mu (x)j_\nu (y) A^a_\lambda (z) |0\rangle ,
\en
satisfy 
\be
q^\lambda T^a_{\mu\nu\lambda}= \int d^4xd^4yd^4ze^{i(k_1\cdot x+k_2\cdot y
-q\cdot z)}\langle 0|Tj_\mu (x)j_\nu (y) \partial^\lambda_z A^a_\lambda (z) 
|0\rangle . 
\en
Using relations (\ref{divfis}) we obtain
\be
\begin{array}{c}
q^\lambda T^{\rm ns}_{\mu\nu\lambda}(k_1,k_2,q)= f_{\rm ns}\left( 
\frac{m^2_\eta\cos\phi_P}{-q^2+m^2_\eta }\Gamma^\eta_{\mu\nu}  (k_1,k_2) + 
\frac{m^2_{\eta^\prime}\sin\phi_P}{-q^2+m^2_{\eta^\prime}}
\Gamma^{\eta^\prime}_{\mu\nu} (k_1,k_2) \right)\\
 - D^{\rm ns}\xi_{\mu\nu}(k_1,k_2).\\ 
q^\lambda T^{\rm s}_{\mu\nu\lambda}(k_1,k_2,q)= f_{\rm s}\left( 
-\frac{m^2_\eta\sin\phi_P}{-q^2+m^2_\eta }\Gamma^\eta_{\mu\nu}  (k_1,k_2) + 
\frac{m^2_{\eta^\prime}\cos\phi_P}{-q^2+m^2_{\eta^\prime}}
\Gamma^{\eta^\prime}_{\mu\nu} (k_1,k_2) \right)\\
 - D^{\rm s}\xi_{\mu\nu}(k_1,k_2).
\end{array} \label{divgammas}
\en
\noi where
\be
\begin{array}{c}
\Gamma^{P}_{\mu\nu} (k_1,k_2)\equiv i\int d^4xd^4yd^4z e^{i(k_1\cdot x+k_2\cdot y
-q\cdot z)}\langle 0|Tj_\mu (x)j_\nu (y) P(z) |0\rangle ,\\
\xi_{\mu\nu}(k_1,k_2)\equiv i\int d^4xd^4yd^4z e^{i(k_1\cdot x+k_2\cdot y
-q\cdot z)}\langle 0|Tj_\mu (x)j_\nu (y) \xi (z) |0\rangle .
\end{array}
\en
In the limit $q\to 0$ the l.h.s in Eqs.(\ref{divgammas}) vanishes yielding the 
following relations
\be
\left( \begin{array}{cc}
f_{\rm ns}\cos\phi_P & f_{\rm ns}\sin\phi_P \\
-f_{\rm s}\sin\phi_P & f_{\rm s}\cos\phi_P
\end{array}\right) 
\left( \begin{array}{c}
\Gamma^{\eta}_{\mu\nu} (k_1,k_2) \\ 
\Gamma^{\eta^\prime}_{\mu\nu} (k_1,k_2)
\end{array}\right) =  
\left( \begin{array}{c}
D^{\rm ns}\xi_{\mu\nu} (k_1,k_2)  \\ 
D^{\rm s}\xi_{\mu\nu} (k_1,k_2)
\end{array}\right) \label{vertices}
\en

Using these results we obtain the invariant matrix element for  
neutral pseudoscalars decaying into two photons 
\be
{\mathcal M}( P^0(q,\eta)\to\gamma(k_1,\epsilon_1)~~\gamma(k_2,\epsilon_2) )
= M(P^0\to\gamma\gamma) \varepsilon (\epsilon_1,k_1,\epsilon_2, k_2).
\en
\noi where
\be 
\begin{array}{c}
M( \pi^0\to\gamma\gamma )= \frac{\alpha}{\pi f_\pi} \\
 M( \eta\to\gamma\gamma )= \frac{\alpha}{3\pi}\left( 
\frac{5}{f_{\rm ns}} \cos\phi_P -\frac{\sqrt{2}}{f_{\rm s}}\sin\phi_P \right)
\\
M( \eta'\to\gamma\gamma )= \frac{\alpha}{3\pi}\left( 
\frac{5}{f_{\rm ns}} \sin\phi_P +\frac{\sqrt{2}}{f_{\rm s}}\cos\phi_P\right).
\end{array} \label{mpgg}
\en

It is worth remarking that the only effect of the $U_A(1)$ anomaly in these processes  
concerns the formation of the physical states $\eta$ and $\eta^\prime $ from 
the flavor
states $\eta^{\rm ns}, \eta^{\rm s}$. This is reflected in the $\sin\phi_P$ and 
$\cos\phi_P$ factors appearing in Eqs.(\ref{mpgg}). Additional effects of 't Hooft  
interaction can be expected in the formation of final states. These effects can be 
important in the case of hadronic final states. For photonic final states the  
$\bar qq-\gamma\gamma $  instanton induced interaction is highly suppressed . We must 
also be clear that the results in Eqs.(\ref{mpgg}) are valid in the soft limit 
$q\to 0$. Extrapolation to the physical $q^2=m^2_\eta $ and $q^2=m^2_{\eta^{\prime}}$ 
are  necessary to compare with existing experimental data. Finally, notice that in the 
soft limit all the information on pseudoscalar meson masses cancels out. Thus the 
extraction 
of the mixing angle using photonic decays of pseudoscalar mesons is free of the 
uncertainties attributed to mechanisms for generation of  meson masses at this 
point.

From  Eqs.(\ref{mpgg}) we obtain the following fractions
\bq\nonumber
\frac{M( \eta\to\gamma\gamma )}{M( \pi^0\to\gamma\gamma )}&=&
\frac{1}{3}\left(5\cos \phi_P-
\frac{\sqrt{2}f_{\rm ns}}{f_{\rm s}}\sin\phi_P\right),\\
\frac{M( \eta'\to\gamma\gamma )}{M( \pi^0\to\gamma\gamma )}&=&
\frac{1}{3}\left(5\sin \phi_P+
\frac{\sqrt{2}f_{\rm ns}}{f_{\rm s}}\cos\phi_P\right) \label{results}
\eq

These results are usually written in terms of
the mixing angle in the singlet-octet basis $\theta_P= \phi_P-\phi_{\rm id}$ 
where$\phi_{\rm id}= 54.7^\circ$ stands for the ideal mixing angle: 
$\cos \phi_{\rm id}=\sqrt{1/3}$, $\sin\phi_{\rm id}=\sqrt{2/3}$. In terms of 
this angle our results read
\bq\nonumber
\frac{M( \eta\to\gamma\gamma )}{M( \pi^0\to\gamma\gamma )}&=&
\frac{1}{\sqrt{3}}\left(\frac{5-2y}{3}\cos \theta_P-
\frac{5+y}{3}\frac{\sqrt{2}}{3}\sin\theta_P\right),\\
\frac{M( \eta'\to\gamma\gamma )}{M( \pi^0\to\gamma\gamma )}&=&
\frac{1}{3}\sqrt{\frac{2}{3}}(5+y)\left(\cos \theta_P+\frac{5-2y}
{\sqrt{2}(5+y)}\sin\theta_P\right),
\eq
\noi Notice that in the case  $f_{\rm ns}=f_{\rm s}$ we recover results from 
$SU(3)$ symmetry or quark model considerations \cite{Groom:2000in}.

The reported data \cite{Groom:2000in} for these fractions is 
\bq
\frac{M( \eta\to\gamma\gamma )}{M( \pi^0\to\gamma\gamma )}&=&
\frac{1.73\pm 0.18}{\sqrt{3}} \\ 
\frac{M( \eta'\to\gamma\gamma )}{M( \pi^0\to\gamma\gamma )}&=&
2\sqrt{\frac{2}{3}}(0.78\pm 0.04) .
\eq

Experimental data  for $\eta\to\gamma\gamma$ decay constrains the mixing angle 
to the range $\phi_P\in \lbrack 38.4^\circ, 47.2^\circ \rbrack$ whereas data 
on $\eta^\prime\to\gamma\gamma$ restrains the same angle to the interval 
$\lbrack 34.3^\circ,41.0^\circ \rbrack$. In the whole, both decays yield

\be
\phi_P\in\lbrack 38.4^\circ, 41.0^\circ\rbrack .
\en
\noi This result is consistent with the averaged value obtained in 
\cite{Feldmann:2000uf,Feldmann:1998vh}. Indeed, our analytical results in 
Eq.(\ref{mpgg}) agree with the corresponding amplitudes in Eqs.(3.13) of 
\cite{Feldmann:1998vh}.
Although both approaches share the virtue of using the flavor basis instead 
of the widely used singlet-octet basis, there are important differences in 
the way this result has been derived. In \cite{Feldmann:1998vh} relations 
(\ref{frels1}) and the generalized PCAC 
relations (see \cite{Feldmann:1997vc}, Eq.(4.2))
\bq \nonumber
\partial^\mu A^{\rm ns}_{\mu} &=& f^{\rm ns}_{\eta}m^2_{\eta}\eta + 
f^{\rm ns}_{\eta^\prime}m^2_{\eta^\prime} \eta^\prime\\ 
\partial^\mu A^{\rm s}_{\mu} &=& f^{\rm s}_{\eta}m^2_{\eta}\eta + 
f^{\rm s}_{\eta^\prime}m^2_{\eta^\prime}\eta^\prime \label{modpcac}
\eq
\noi have been assumed and used to calculate  
the  two photon decay widths of neutral pseudoscalars. In the present work, we 
start with 
a chiral Lagrangian which incorporates $U_A(1)$ symmetry breaking in a way 
inspired by instanton calculations. As a result we obtain a dynamical 
mechanism for the (OZI rule violating) mixing of pseudoscalar strange and 
non-strange quarkonia, namely, the coupling of the $U_A(1)$ symmetry breaking 
to the v.e.v's of scalars due to the spontaneous breakdown of chiral symmetry. 
In this framework we are able to calculate modifications  to the naive
 PCAC relations due to the $U_A(1)$ anomaly. In general, these modifications 
involve linear, bilinear and trilinear combinations of meson fields. However,  
only linear terms contribute to  two photon decay of pseudoscalars at tree 
level and for this particular case we derive relations 
(\ref{frels1}) and the modified PCAC relations (\ref{modpcac}) (first term 
in the r.h.s. of Eq.(\ref{divfis})) which are the starting point in 
\cite{Feldmann:1998vh}.

\bi

\section{Conclusions}

\bi

We use a $U(3)\times U(3)$ effective chiral model incorporating $U_A(1)$ symmetry 
breaking to study  the strong and electromagnetic contributions to the 
non-conservation of the isosinglet axial currents in flavor basis. Strong 
contributions are explicitly calculated in the model and related to the mass 
matrix of isoscalar pseudoscalr fields and the corresponding weak decay constants. 
Electromagnetic contributions are introduced using 
't Hooft's argument on anomaly matching. We calculate the $P^0\to\gamma\gamma$ decays 
in this framework and show that the only effect of the strong anomaly in these 
processes is in the formation of the $\eta$ and $\eta^\prime$ systems. We use these 
results to estimate the pseudoscalar mixing angle from  experimental data on 
$\eta\to\gamma\gamma$ and $\eta^\prime\to\gamma\gamma$ obtaining 
$\phi_P\in\lbrack 38.4^\circ, 41.0^\circ\rbrack$

\end{document}